\documentclass[11pt]{elsarticle}
\biboptions{sort&compress,numbers}
\usepackage{microtype}
\usepackage{algorithm}
\usepackage{algpseudocode}
\usepackage[margin=1in]{geometry}
\usepackage{amsmath,amssymb,amsthm,bm,stmaryrd}
\usepackage{subcaption}
\biboptions{sort&compress}
\usepackage{upgreek}
\usepackage{epstopdf}
\usepackage{float}
\usepackage{array}
\usepackage{xcolor}
\usepackage{gensymb}
\usepackage[labelfont=bf]{caption}
\usepackage[export]{adjustbox}
\AppendGraphicsExtensions{.tif}

\usepackage[final]{changes} 
\definecolor{C0}{HTML}{1F77B4}
\definecolor{C1}{HTML}{FF7F0E}
\definecolor{C2}{HTML}{2ca02c}
\definecolor{C3}{HTML}{d62728}
\definecolor{C4}{HTML}{9467bd}
\definecolor{C5}{HTML}{8c564b}
\definechangesauthor[color=C0]{R1} 
\definechangesauthor[color=C1]{R2} 
\definechangesauthor[color=C3]{RA} 

\begin{document}


    \begin{frontmatter}

        \title{Ductile fracture in functionally graded materials: Insight into crack behavior within the gradient interface}
        \author[cmu]{Katherine Piper}
        \ead{kbpiper@andrew.cmu.edu}
        \author[lanl]{Vinamra Agrawal\corref{cor1}}
        \ead{vagrawal@lanl.gov}
        \cortext[cor1]{Corresponding author}
        \address[cmu]{Department of Civil and Environmental Engineering, Carnegie Mellon University, Pittsburgh PA, USA}
        \address[lanl]{XCP-5, Los Alamos National Laboratory, USA}
        
        \begin{abstract}
        Advances in manufacturing has enabled fabrication of metallic functionally graded materials (FGMs) for various industrial applications. 
        While brittle fracture behavior of FGMs have received attention from the community, the role of plasticity and overall ductile fracture behavior of metallic FGMs still needs to be studied. 
        In this work we study the crack propagation in ductile FGMs, specifically focusing on crack propagation within the gradient region of an FGM. 
        We  investigate the effects of plasticity in crack propagation and the study correlations between accumulated plastic strain and crack growth patterns in an FGM. 
        Through this, we determine key differences in crack growth patterns between well-studied brittle FGMs, and more recently developed ductile FGMs. 
        We provide insights into the influence of both the angle of incidence and the width of the gradient, and show the importance of plasticity on crack growth within an FGM. 
        \end{abstract}

    
    \end{frontmatter}
    
    \section{Introduction} \label{Sec: Introduction}
        Functionally graded materials (FGMs) have had emerging interest in recent years. 
        In functionally graded materials, a single bulk material contains a variational composition. 
        Frequently, this consists of two materials, with a gradient in composition from one material to the other. 
        Specific gradient materials have a variety of use cases, from nanotechnology to materials with multiple magnetic properties for use in electric motors \cite{JI2023101194}. 
        When two materials are bonded, diffusion between the materials naturally leads to some gradation in properties in the joint. 
        However, in FGMs, this graded region is intentionally engineered to improve properties between the materials. 
        Manufacturing methods for FGMs vary, from the addition of various ratios of glass fiber to epoxies, to additive manufacturing \cite{LU200438}. 
        Functionally graded additively manufactured alloys, such as a gradation between Hiperco50a and stainless steel, have potential for massive improvement in thermal performance of electric motors and ion engines, by reducing part count and shortening the thermal path out of motor cores \cite{Hoffmann}. 
        Other applications for additively manufactured ductile FGMs include the gradient manufacturing between nickel alloys and copper, with uses in jet and rocket propulsion systems \cite{GRANDHI202347}. 
        Ductile functionally graded materials are frequently fabricated through directed energy deposition (DED) \cite{KELLY2021101845, SVETLIZKY2021271, SCHNEIDERMAUNOURY201755, Balkan}.
        DED is a form of additive manufacturing, which feeds material in front of a high power laser, fusing the material together. 
        This manufacturing method can also be used to create gradient alloys. 
        By feeding two different materials in front of the laser simultaneously, new gradient alloys are formed \cite{GRANDHI202347,Ahn,Ahn2,LOH201834,coatings13040773}.
        Significant research effort has been used for determining which materials can be manufactured through this method. 
        Factors such as lattice structure and bonding play a significant role in the ability to be graded, and material strength within the graded regions can vary substantially \cite{Greaves}.
        
        While research has continued on manufacturing of graded alloys, there have been few studies on the mechanical behavior under high strain rates of FGMs. 
        Before FGMs are implemented for widespread use, including military and critical NASA applications, a detailed computational understanding of failure modes in FGMs is needed. 
        Certification of these components remains a challenge, as additively-manufactured mission-critical components must meet the same standards of traditional subtractive manufacturing processes \cite {BLAKEYMILNER2021110008}.
        Phase field (PF) modeling has been shown to robust at predicting crack growth under a wide range of conditions \cite{Kristensen, BORDEN2016130, MIEHE20161}, initially within brittle materials, and later extended to ductile materials \cite{Ambati,SAMANIEGO2021106424,MIEHE2015486,MIEHE2016619,Ambati2}.
        Operating using the equilibrium definition of fracture \cite{IV_Griffith} the variational form of phase field modeling builds on Griffith's crack theory by minimizes the potential energy function. 
        The brittle form of PF fracture proposes modifications to Griffith's theory, increasing stability and accuracy of the simulation \cite{FRANCFORT19981319}.
        The stability afforded by PF modeling is advantageous for complex geometries and crack growth cases \cite{Agrawal}.
        
        In FGMs, the subcomponents of the material, and thus the overall material properties, vary across the domain of the material. 
        Properties such as $K_{1C}$, elastic modulus $E$, and Poisson's ratio $\nu$ have a strong relationship with material \cite{Greaves}. 
        In brittle fracture, this has been shown to create unique material properties and fracture mechanics behaviors\cite{HIRSHIKESH2019239, ARAVIND2023109344, PANG2024110286, shao2021adaptive}. 
        Modeling these materials using the PF method requires specific modeling advances, including creation of a sufficiently fine mesh in the region of the crack\cite{Kim}. 
        A highly refined mesh, however, increases computational time exponentially with each level of refinement. 
        Using block structured adaptive mesh refinement, the mesh can be refined in targeted regions where high refinement is necessary. 
        Block structured adaptive mesh refinement allows for increased parallelization in solving, and avoids communication overloads from other mesh refinement methods \cite{RUNNELS2021110065}. 
        The next challenge of using phase field methods is due to the sharp interface between a crack and uncracked material.
        Phase field methods rely on a set of continuously differentiable material properties, while cracked interfaces are a discontinuity. 
        This is efficiently addressed by solving the phase field equations with the smooth boundary method, circumventing the challenges caused by meshing discrete interfaces \cite{Agrawal}. 
        
        In addition to the challenges of modeling phase field fracture, modeling FGMs present an additional challenge. 
        Existing phase field fracture models use either Mode I tension testing, or simulated 4-point bending. 
        Convergence of the model is dependent on several factors, including the relation between the resultant force from the applied displacement and the fracture strength of the material. 
        In cases where the two graded materials have substantially differing properties, the model can diverge as the crack passes into the region of the weaker material. 
        However, DED of FGMs is primarily used between similar alloys, including between nickel and steel alloys \cite{CARROLL201646}, titanium alloys \cite{LI201716638}, and aluminum alloys \cite{app8071113}.
        
        While recent studies in molecular dynamics have better informed engineering design with gradient alloys \cite{GAO2023258,Mitra}, these studies provide little insight on fracture and other end-of-life phenomena. 
        Recent advances in phase field ductile fracture models \cite{Agrawal, RUNNELS2021110065} provide the basis for new insights into these phenomena. While some work has been conducted on ductile phase field fracture \cite{AZINPOUR2023103906}, we specifically aim to study crack growth within a graded region.
        
        In this work, we present the implementation and preliminary verification of ductility into a phase field model which allows for efficient and accurate computation of fracture in metal FGMs. 
        The paper is structured as follows. Section \ref{Sec:Methodology} is divided into 2 sections. 
        (1) We discuss the phase field formulation of ductile fracture, and (2) the numerical implementation of J2 Plasticity in a near-singular finite difference solver. 
        Following, in Section \ref{Sec: Problem Formulation}, we discuss the problem setup, materials selection, and interfaces. 
        Following, in Section \ref{Sec:Results}, we show the importance of incorporating ductility into phase field models for functionally graded materials, and then present several example simulations of cracks passing through functionally graded interfaces. We use the results of this to provide an estimate of three differing fracture regimes in functionally graded materials.
        
    \section{Methodology} \label{Sec:Methodology}
        \subsection{Phase field formulation for ductile fracture}
            In phase field methodology, the sharp crack discontinuity is replaced by a regularized smooth field $c(\bm{x}) \in [0,1]$ with a characteristic length scale $\xi$.
            Conventionally, for a brittle material, the energy functional $\mathcal{L}$ is formulated as 
            \begin{equation}
                \mathcal{L}\left(\bm{u},c\right) = \int_\Omega \left(g(c) + \eta\right) W\left(\bm{\varepsilon}(\bm{u})\right) d\bm{x} + \int_\Omega G_c\left[ \frac{w(c)}{4\xi} + \xi |\nabla c|^2 \right] d\bm{x},
            \end{equation}
            where $W(\bm{\varepsilon}(\bm{u}))$ is the strain energy density of the material with strain $\bm{\varepsilon} = \text{sym} (\nabla \bm{u})$, $G_c$ is the fracture energy, $g(c)$ and $w(c)$ are degradation and geometric functions, and $\eta$ is a small parameter introduced for computational stability.
            The degradation function $g(c)$ takes the value of 0 inside the crack and 1 outside, while the geometric function takes the value 1 inside the crack and 0 outside.
            We account for the tension-compression asymmetry by splitting the strain energy density into positive and negative parts $W = W^+ + W^-$ using the spectral decomposition $\bm{\varepsilon} = \sum_{i=1}^3 \varepsilon_i \hat{\bm{v}}_i \otimes \hat{\bm{v}}_i$ of strain,
            \begin{equation}
                W^\pm = W(\bm{\varepsilon}^\pm), \quad \bm{\varepsilon}^\pm = \sum_{i=1}^d \varepsilon_i^\pm \, \hat{\bm{v}}_i \otimes \hat{\bm{v}}_i.
            \end{equation}
            We modify the energy functional for brittle fracture as
            \begin{equation}
                \mathcal{L}_{\text{brittle}} \left(\bm{u},c\right) = \int_\Omega \left[\left(g(c) + \eta\right) W^+ + W^- \right] d\bm{x} + \int_\Omega G_c\left[ \frac{w(c)}{4\xi} + \xi |\nabla c|^2 \right] d\bm{x},
            \end{equation}

            We use an isotropic elasto-plastic material with linear hardening in this work. 
            The strain energy density is given by
            \begin{equation}
                W\left(\varepsilon_e\right) = \frac{1}{2} \bm{\varepsilon}_e:\mathbb{C}\bm{\varepsilon}_e\quad \text{where} \quad \bm{\varepsilon}_e = \bm{\varepsilon} - \bm{\varepsilon}_p
            \end{equation}
            The elastic modulus tensor $\mathbb{C}$ of the isotropic material is given by
            \begin{equation}
                \mathbb{C}_{ijkl} = \lambda\, \delta_{ij}\delta_{kl} + \mu \left(\delta_{il}\delta_{jk} + \delta_{ik}\delta_{jl} \right),
            \end{equation}
            where $\lambda$ and $\mu$ are Lam\'{e} constants. 
            We model the evolution of plastic strain with $J_2$ plasticity model with linear hardening behavior. 
            The yield strength $K(\alpha)$ is given by
            \begin{equation}
                K(\alpha) = \sigma_Y + \theta \bar{H} \alpha, \quad \theta\in [0,1],
            \end{equation}
            where $\sigma_Y$ is the yield strength, $\bar{H}$ is the hardening modulus, $\theta$ is a material parameter (chosen as $1$ in this work), and $\alpha$ is the accumulated plastic strain.
            We update the energy functional for ductile failure with plastic energy as
            \begin{align}
                \mathcal{L}_{\text{ductile}} (\bm{u}, \alpha, c) = &\int_\Omega \left[\left(g(c) + \eta\right) W^+(\bm{\varepsilon}_e) + W^-(\bm{\varepsilon}_e) \right] d\bm{x} + \int_\Omega G_c\left[ \frac{w(c)}{4\xi} + \xi |\nabla c|^2 \right] d\bm{x} \nonumber \\
                &+\int_\Omega\left(g(c) + \eta\right) \left( \sigma_Y + \frac{1}{2} H \alpha \right) \alpha\, d\bm{x} \label{eq:pf_ductile_energy}
            \end{align}
        \subsection{Numerical implementation}
            In this section, we describe the numerical implementation for solving for stationarity of the energy functional $\mathcal{L}_{\text{ductile}}$ to obtain the equilibrium solution.
            \subsubsection{Near singular finite difference solver}
                We use Alamo, an open-source finite-difference-based multilevel, multigrid and multicomponent solver capable of handling near singular problems.
                Alamo uses AMReX libraries for highly scalable data structures and block-structured adaptive mesh refinement (BSAMR) capabilities.
                Alamo uses smooth boundary element method (SBEM) to regularize sharp interfaces and material boundaries.
                Alamo has been used for a variety of solid mechanics and multiphysics problems, such as microstructure evolution \cite{runnels2020phase}, fracture \cite{agrawal2021block}, topology optimization \cite{Agrawal}, and deflagaration \cite{meier2024finite}.

                We use Alamo to solve elastic equilibrium, crack equilibrium, and plasticity update in a staggered fashion. 
                We use Alamo to implicitly solve elastic equilibrium equation in strong form
                \begin{align}
                    &\text{div}\;\; \bm{\sigma} = \bm{0} \;\;\; \text{ on }\Omega,\qquad \bm{\sigma} = \mathbb{C} \left(\bm{\varepsilon} - \bm{\varepsilon}_p\right) \nonumber \\
                    &\bm{\sigma}\hat{\bm{n}} = \bm{t}\;\;\;\text{ on }\partial\Omega
                \end{align}
                Next, we perform plasticity update using the radial return algorithm described in Section \ref{sec:RadialReturn}.
                
                The stationarity condition of the energy function $\mathcal{L}_{ductile}$ for crack equilibrium yields
                \begin{equation}
                    0 = g'(c) W^+(\bm{\varepsilon}_e) - G_c \left[\frac{w'(c)}{2\xi} + 2\xi \Delta c \right] + g'(c) \left( \sigma_Y + \frac{1}{2} H \alpha \right) \alpha
                \end{equation}
                In this work, we replace the above equation with a Ginzburg-Landau type evolution law
                \begin{equation}
                    \dot{c} = -M \left[ g'(c) \mathcal{H}^+ - G_c \left[\frac{w'(c)}{2\xi} + 2\xi \Delta c \right] + g'(c) \left( \sigma_Y + \frac{1}{2} H \alpha \right) \alpha \right], \label{eq:GinzburgLandau}
                \end{equation}
                where $M\geq 0$ is the crack mobility and $\mathcal{H}^+$ is introduced for crack irreversibility as 
                \begin{equation}
                    \mathcal{H}^+ := \max_{\tau\in[0,t]} W^+(\bm{\varepsilon}_e)
                \end{equation}
                We use equation (\ref{eq:GinzburgLandau}) to evolve the crack field until equilibrium $\dot{c} =0$ using explicit integration techniques. 
                Through this, we emphasize that equation (\ref{eq:GinzburgLandau}) is merely used as a substitute for gradient-descent method and should not be interpreted as a kinetic law. 
                Therefore, the mobility factor $M$ is just used to control the step size so as not to violate the CFL conditions.
                We chose quartic degradation function $g(c)=4c^3-3c^4$ and geometric function $w(c)=1-g(c)$ following \cite{kuhn2015degradation}.
            \subsubsection{Radial return algorithm for J2-plasticity}\label{sec:RadialReturn}
                We use the classic radial-return algorithm for evolving plastic strains.
                We describe the algorithm briefly below.
                For a detailed description, we refer the reader to \cite{simo2006computational}.
                We use the internal variables $\bm{q}=\{\alpha, \bm{\varepsilon}_p, \bm{\beta}\}$ in our implementation, where $\bm{\beta}$ is the center of the von-Mises yield surface.
                The yield condition, flow rule, and the hardening rule are 
                \begin{align}
                    &\bm{\eta}:=\text{dev}[\bm{\sigma}] - \bm{\beta}, \quad \text{tr}\,\bm{\beta} := 0,\quad f(\bm{\sigma},\bm{q}) = ||\bm{\eta}|| - \sqrt{\frac{2}{3}} K(\alpha), \nonumber \\
                    &\dot{\bm{\varepsilon}}_p = \gamma \frac{\bm{\eta}}{||\bm{\eta}||}, \quad\dot{\alpha} = \gamma \sqrt{\frac{2}{3}},\quad \dot{\bm{\beta}} = \frac{2}{3}\gamma(1-\theta) \bar{H} \frac{\bm{\eta}}{||\bm{\eta}||} \label{eq:J2Plasticity}
                \end{align}

                We implement the radial return algorithm to solve equation (\ref{eq:J2Plasticity}) and obtain the updated internal variables $\bm{q}$.
                Given the stress $\bm{\sigma}_n$ and strain $\bm{\varepsilon}_n$ at time $t_n$, and strain $\bm{\varepsilon}_{n+1}$ at the next time step, we obtain the updated internal variables through the process outlined in Algorithm \ref{alg:J2RadialReturn}.
                \begin{algorithm}
                    \caption{Radial return algorithm for J2 plasticity}
                    \label{alg:J2RadialReturn}
                    \begin{algorithmic}[1]
                        \State $\bm{e}_n \gets \text{dev}\,\bm{\varepsilon}_n, \;\bm{e}_{n+1} \gets \text{dev}\,\bm{\varepsilon}_{n+1},\;\bm{s}_n \gets \text{dev}\bm{\sigma}_n$ \Comment{compute deviatoric quantities}
                        \State $\bm{s}_{n+1}^{trial} \gets \bm{s}_n + 2\mu (\bm{e}_{n+1}-\bm{e}_n),\; \bm{\eta}_{n+1}^{trial} \gets \bm{s}_{n+1}^{trial} - \bm{\eta}_n$ \Comment{compute trial quantities}
                        \State $\bm{n}_{n+1} = \bm{\eta}^{trial}_{n+1} / ||\bm{\eta}_{n+1}^{trial} ||$ \Comment{compute new yield surface normal}
                        \State $\text{Solve}:\; ||\bm{\eta}_{n+1}^{trial}|| - \sqrt{\frac{2}{3}} K(\alpha_n + \sqrt{\frac{2}{3}} \Delta \gamma) \equiv 0$ \Comment{compute incremental plastic strain}
                        \State $\alpha_{n+1}=\alpha_n+\sqrt{\frac{2}{3}} \Delta \gamma$  \Comment{update internal variable $\alpha$}
                        \State $\bm{\beta}_{n+1} = \bm{\beta}_n + \sqrt{\frac{2}{3}} \theta \bar{H} (\alpha_{n+1}-\alpha_n) \bm{n}_{n+1}$ \Comment{update internal variable $\bm{\beta}$}
                        \State $\bm{\varepsilon}^p_{n+1} = \bm{\varepsilon}^p_n + \Delta\gamma\, \bm{n}_{n+1}$\Comment{update plastic strain $\bm{\varepsilon}_p$}
                    \end{algorithmic}
                \end{algorithm}
            
    \section{Problem setup} \label{Sec: Problem Formulation}
\begin{figure}[hbt!]
        \centering
          \includegraphics[width=.6\linewidth]{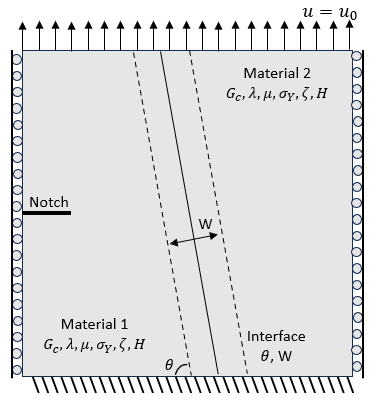}
          \caption{Parameter Space in a Phase-Field Fracture study of a functionally graded ductile material}
        \label{fig:ParameterSpace}
        \end{figure}

        We study the crack propagation of a ductile FGM in Mode-I configuration as shown in Figure \ref{fig:ParameterSpace}.
        Unless otherwise stated, our simulations consist of a square domain $\bm{x}\in [-L,L]\times [-L,L]$ with a base mesh of $64\times 64$ cells and six BSAMR levels.
        For computational stability, we choose $\eta = 10^{-4}$ in equation (\ref{eq:pf_ductile_energy}).
        We represent each material with a material variable $\phi_i \in [0,1]$ and the crack field with $c \in [0,1]$.
        We initialize our domain with using an error function with a length scale $W$ such as $\phi_1$ ($\phi_2$) takes the value zero (one) in the region occupied by Material 2, one (zero) in the region occupied by Material 1, and smoothly transitions over $W$.
        We introduce a horizontal notch of length $L/2$ at the center of left edge.
        We then apply boundary conditions as shown in Figure \ref{fig:ParameterSpace} and impose a fixed vertical displacement $\bm{u}_0$ on the top edge to create a Mode-I loading scenario. 
        
        We choose our material properties ($\lambda$, $\mu$, $G_c$, $\sigma_Y$, $H$, and $\xi$) to correspond to two Aluminum alloys. 
        We selected these materials for two reasons: 1) high variance in ductility, and 2) crystal lattice compatibility which is a critical property for manufacturing real gradients \cite{Hoffmann}.
        We choose the material property values as $\lambda_1 = 52.3242\times 10^9$, $\mu_1 = 26.9549 \times 10^9$, $\sigma_{Y1} = 503\times 10^6$, $H_1 = 15.9 \times 10^9$, $G_{c1} = 7.766 \times 10^3$, and $\xi_1=1.0\times10^{-5}$ for Material 1, and $\lambda_2 = 50.2808\times 10^9$, $\mu_2=25.9023\times 10^9$, $\sigma_{Y2} = 41.4 \times 10^6$, $H_2= 15.9 \times 10^9$, $G_{c2} = 61.575 \times 10^3$, and $\xi_2=1.0\times10^{-5}$ for Material 2.
        These values are computed from published values for commercial alloys of Aluminum 3003 and Aluminum 7075 at $25\degree C$\cite{10.31399/asm.hb.v02.9781627081627}. 

        Within the interface, we follow a linear interpolation based on the material variables $\phi_i$ to obtain the properties of the interface. 
        Experimental studies have shown a nonlinear variance in mechanical properties of gradient interfaces \cite{aifantis2014gradient}.
        However, such experiments are difficult and there is a need for nonstandard, novel experimental methodologies to accurately quantify the variation of properties through a gradient interface.
        A linear interpolation or weighted average of material properties within the interface is widely used due to ease of implementation and understanding \cite{Mirzaali, AZINPOUR2023103906, ASURVIJAYAKUMAR2021107234}. 
        We reserve the topic of accurately accounting for changes in material properties through the interface for future studies and continue here with the choice of linear interpolation.

        For our simulations, we choose the gradient to be centered in the domain and adjust the interface length scale $W$ and the angle of interface $\theta$ for our parametric study.
        We choose three values of $\theta$ and four values of $W$ for a total of twelve test cases.
        Specifically, we choose $\theta = 18^\circ,\, 45^\circ,\,$ and $72^\circ$ and $W / L$ ratio as $1/6,\, 1/3,\, 1$, and $2$.
        
    \section{Results} \label{Sec:Results}
    In this section we present key findings of our study on fracture in ductile FGMs. 
    First, we highlight the difference between brittle and ductile fracture behavior in metallic FGMs. 
    Next, we study the evolution of plastic strain as the crack propagates and highlight the correlation between plastic strain and crack behavior.
    Finally, we present the results of our parametric study on crack behavior for varying gradient interface design.
    
    \subsection{Impact of Plasticity} \label{Sec:Impact of Plasticity}
\begin{figure}[htb]
        \centering
        \begin{subfigure}{.445\textwidth}
          \centering
          \includegraphics[width=.95\linewidth]{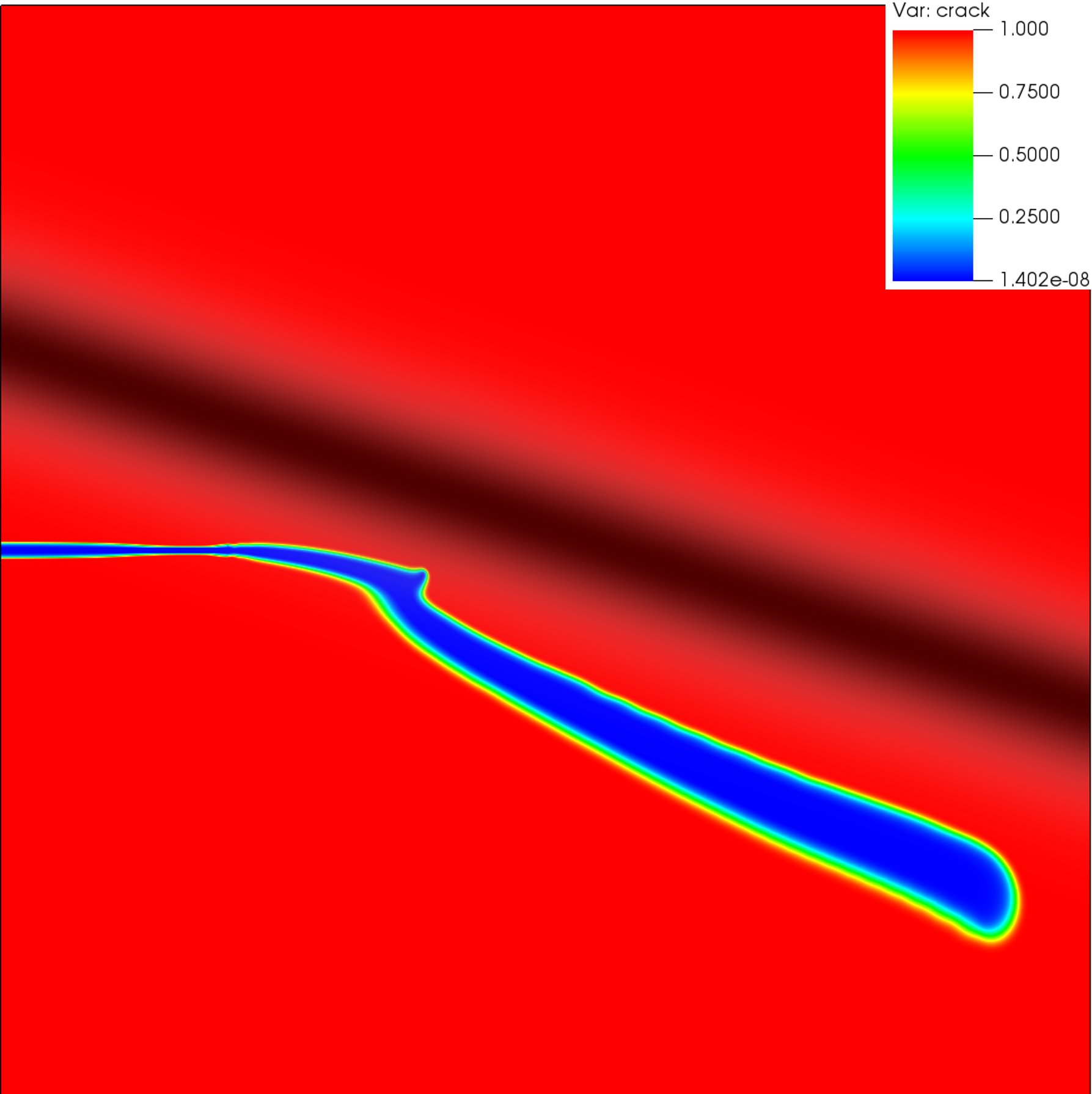}
          \caption{Brittle Case}
          \label{fig:23.lg.brittle}
        \end{subfigure}
        \begin{subfigure}{.45\textwidth}
          \centering
          \includegraphics[width=.95\linewidth]{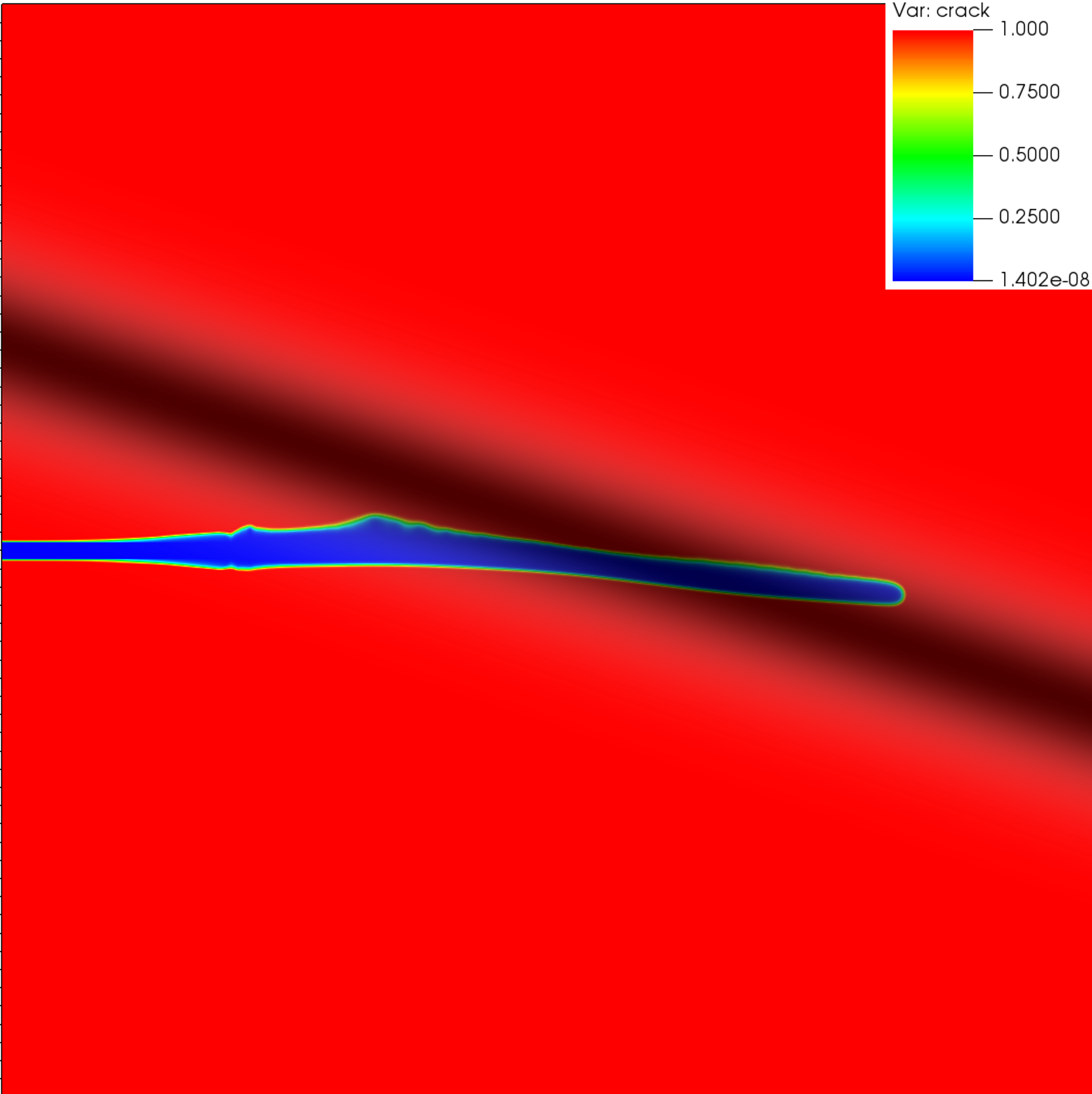}
          \caption{Ductile Case}
          \label{fig:23.lg.ductile}
        \end{subfigure}
        \caption{Variance in phase field crack growth given identical material properties and loading conditions, one case run as a brittle simulation and one case run as a ductile simulation. This simulation is conducted over a square domain of $x\in[-0.06,0.06]\times [-0.06,0.06]$.}
        \label{fig:ductvsbrit}
        \end{figure}

    Substantial previous work has provided valuable insight into the effects of functional gradation on crack behavior in brittle materials \cite{ASURVIJAYAKUMAR2021107234, HIRSHIKESH2019239, ARAVIND2023109344, PANG2024110286}. 
    The phase field formulation has recently been extended to study ductile functionally graded materials \cite{AZINPOUR2023103906, shao2021adaptive}.
    However, the role of gradient architecture and specific mechanical effects on crack propagation has yet to be systematically studied.
    Here we demonstrate the importance of incorporating  ductility in the overall fracture behavior of metallic FGMs.
    While ductile and brittle fracture are characteristically different problems with different underlying microstructural mechanisms, crack growth under certain conditions can be similar.
    Due to ease of computational implementation and similarity in some cases, the role of plasticity is sometimes overlooked or neglected.
    However, plasticity can lead to significantly different behavior as shown in Figure \ref{fig:ductvsbrit}.
    Figure \ref{fig:ductvsbrit} compares brittle and ductile fracture behavior for identical gradient geometries and loading conditions. 
    We note that in the case of brittle (Figure \ref{fig:ductvsbrit}a), the crack deflects from the interface while the ductile crack (Figure \ref{fig:ductvsbrit}b) continues to propagate through the interface into Material 2.
    
\begin{figure}[ht]
          \centering
          \includegraphics[width=.8\linewidth]{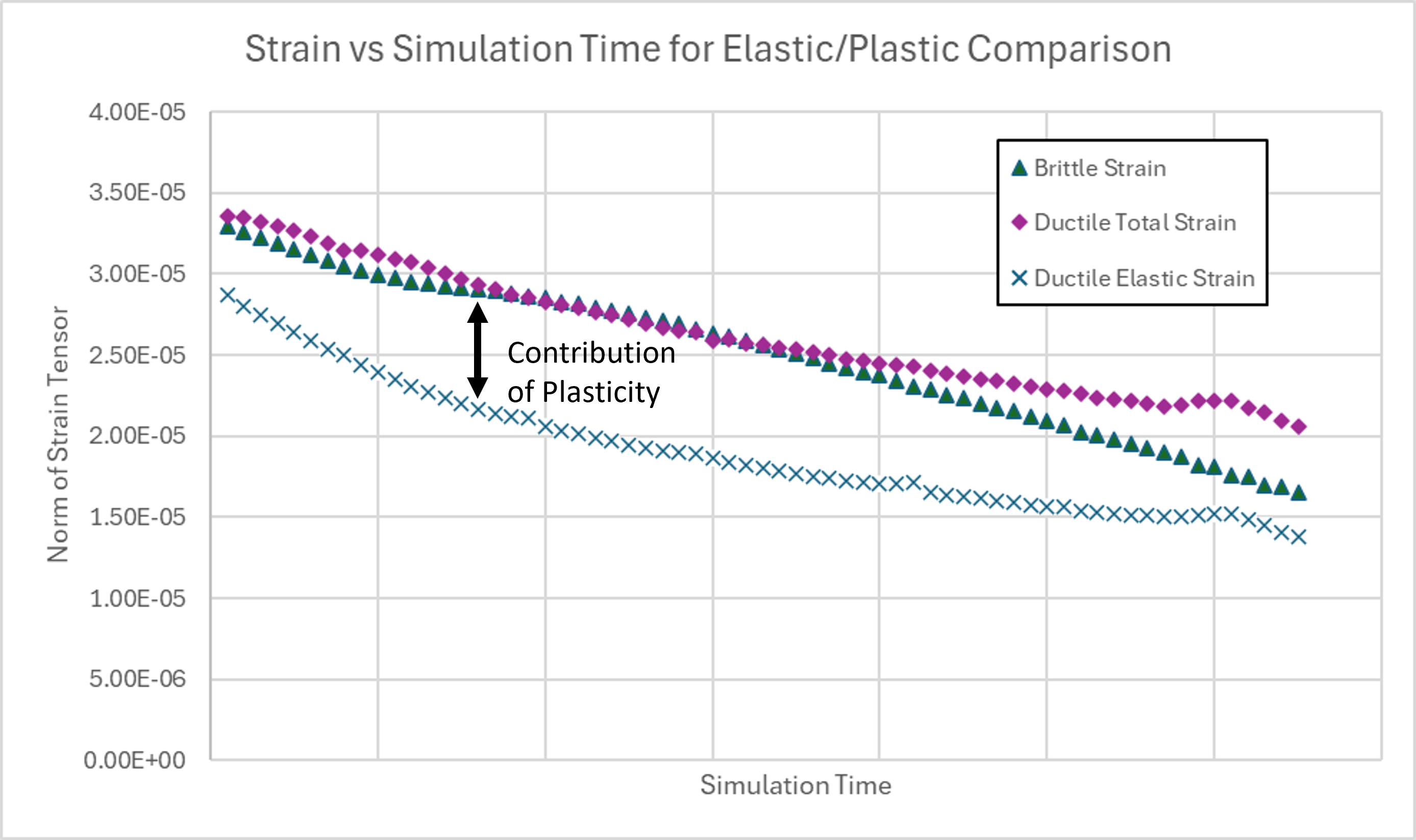}
        \caption{Variance in the norm of the strain tensor between ductile and brittle models for crack propagation, with identical geometries and boundary conditions}
        \label{fig:ductvsbritstrain}
        \end{figure}

    The contribution of plasticity is readily evident in a plot of total strain shown in Figure \ref{fig:ductvsbritstrain}. 
    We note that the plastic strain accumulation makes up roughly 15\% of total strain, indicating that a significant portion of energy dissipation is associated with plastic dissipation, which in turn affects the crack driving forces.
    
    \subsection{Plastic Strain Evolution} \label{Sec:Plastic Strain Evolution}
    In this section, we investigate the plastic strain evolution during crack propagation through the gradient region.
    During our parameter space investigation, discussed in Section \ref{Sec:Impact of Gradation Parameters}, we observed three distinct behaviors of crack: deflection, continued propagation, and arrest.
    These behaviors are clearly defined by a plot of the total plastic strain present throughout the cycle.
    In all simulations, we observe a sharp spike in plasticity at the start of the simulation corresponds to plastic strain buildup near the crack tip, the point at which the initial notch begins to propagate.
    The magnitude of this initial spike in plastic strain is dependent only on initial notch geometry and loading conditions.
    Following this initial spike, plastic strain remains relatively constant, and low, while propagating through the more brittle of materials.
    Upon approaching the gradient, the plastic strain demonstrates one of three behaviors. 
    
\begin{figure}[hbt!]
        \centering
          \includegraphics[width=.6\linewidth]{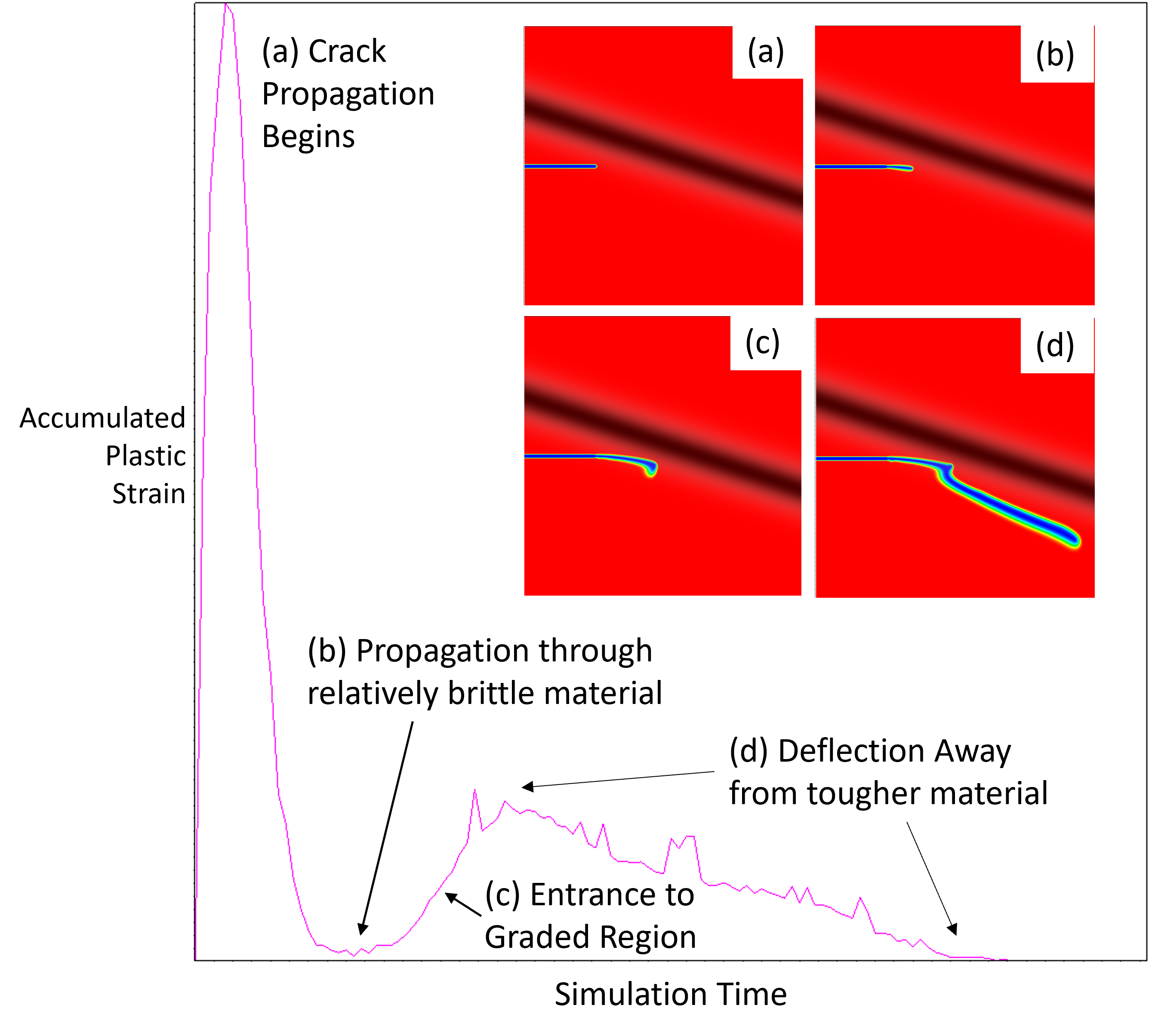}
          \caption{Plastic Strain around a crack which deflects upon approach to a material gradient}
        \label{fig:DeflectionStrain}
    \end{figure}

    First, some cracks deflected away from the graded region, remaining in the material with lower $G_c$. 
    In this case, plasticity increases as the crack approaches the gradient, and then reaches an inflection point, and begins to decrease as the crack moves away from the graded region. 
    This behavior is most commonly observed for shallow angles of incidence from the crack, and relatively narrow gradient regions.
    This case is annotated in Figure \ref{fig:DeflectionStrain}.
    
\begin{figure}[hbt!]
        \centering
          \includegraphics[width=.6\linewidth]{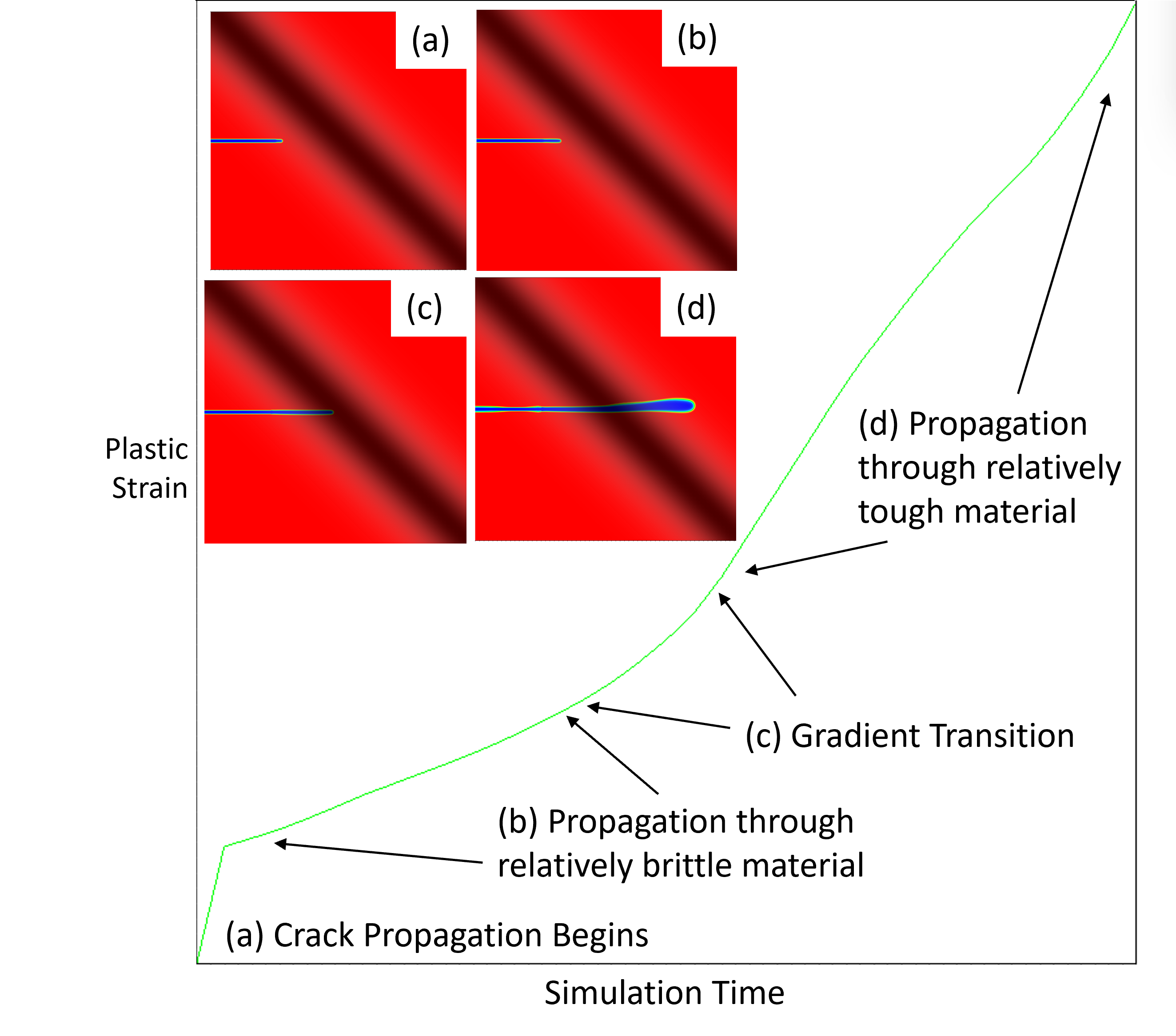}
          \caption{Plastic Strain around a crack which continues to propagate through a material gradient}
        \label{fig:PropagationStrain}
    \end{figure}

    Second, in some cases, it is energetically favorable for the crack to continue propagation through the gradient. 
    In these cases, we see a slight inflection in plastic strain as the crack propagates into the region. 
    As the crack continues propagation into the tougher material, the slope of plastic strain remains higher, corresponding to the higher accumulated strain in a tougher material.
    This behavior is most common with wide gradients and shallow angles of incidence from the crack to the gradient.
    This behavior is shown in Figure \ref{fig:PropagationStrain}.
    
\begin{figure}[hbt!]
        \centering
          \includegraphics[width=.6\linewidth]{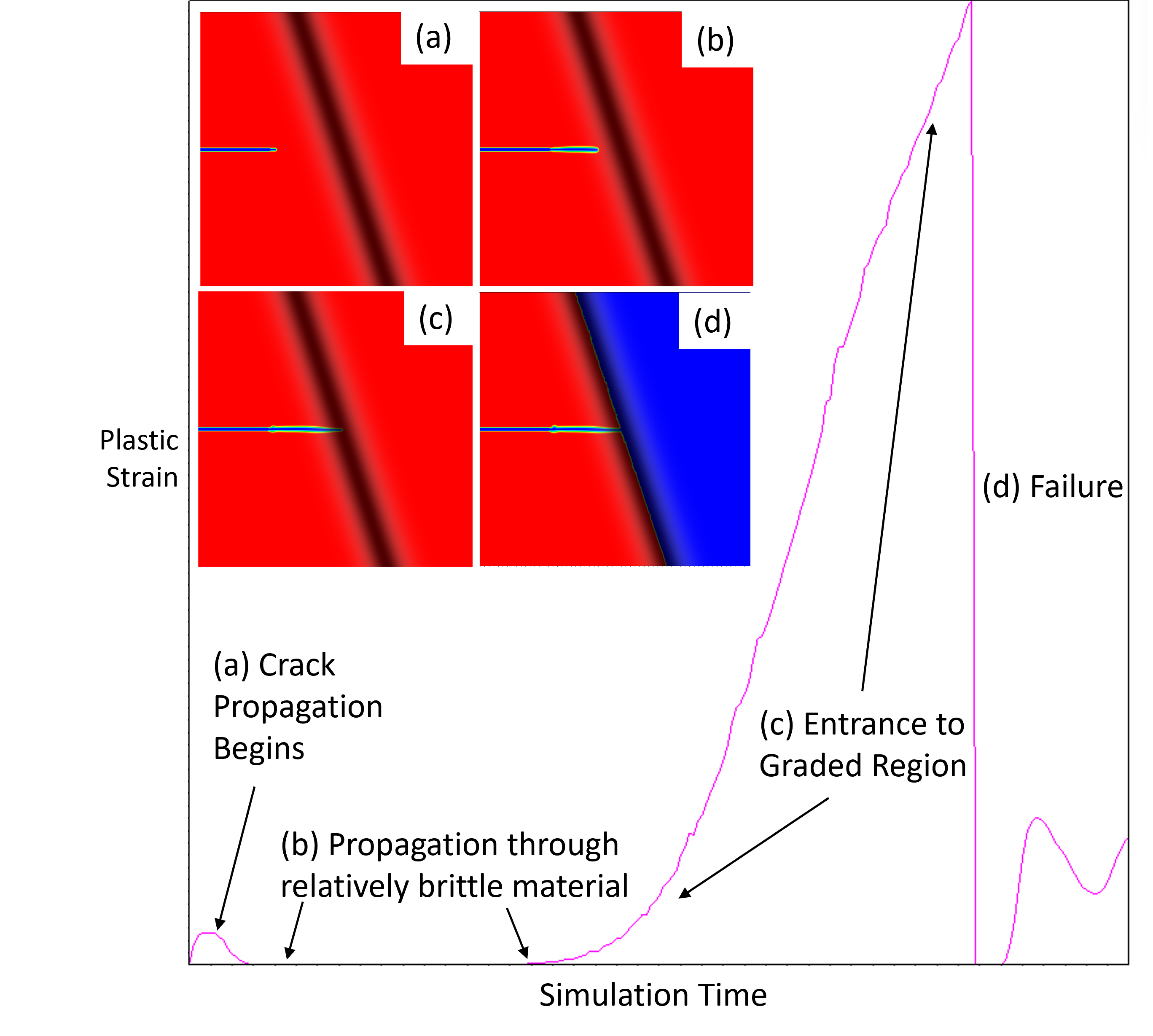}
          \caption{Plastic Strain around a crack which is arrested by a material gradient}
        \label{fig:ArrestingStrain}
    \end{figure}

    Finally, in some cases the gradient results in crack arresting behavior. 
    In a plot of plastic strain, this behavior is visible as an increase in plastic strain as the crack approaches the gradient, followed by a sharp drop-off indicating failure. 
    This behavior is most commonly seen in cases with steep angles of incidence and narrow graded regions.
    This behavior is shown in Figure \ref{fig:ArrestingStrain}.
    \subsection{Parametric Study} \label{Sec:Impact of Gradation Parameters}
    Here, we discuss the results of a parametrization study on the gradation.
    We modified two parameters - the angle $\theta$ with which the crack approaches the gradient and the width of the gradient region $W$. 
    We investigated three angles, $\theta = 18^\circ$ (low), $\theta= 45^\circ$ (medium), and $\theta = 72^\circ$ (high), and four widths of the gradient, $W/L = 1/6,\, 1/3\, 1$, and $2$.
    As mentioned previously, we observed three behaviors in crack propagation: crack deflection, continued propagation, and arrest. 
    Across our parameter space, we observed deflection in two simulations, continued propagation in eight simulations, and arresting behavior in two simulations. 
        
\begin{figure}[hbt!]
        \centering
          \includegraphics[width=.55\linewidth]{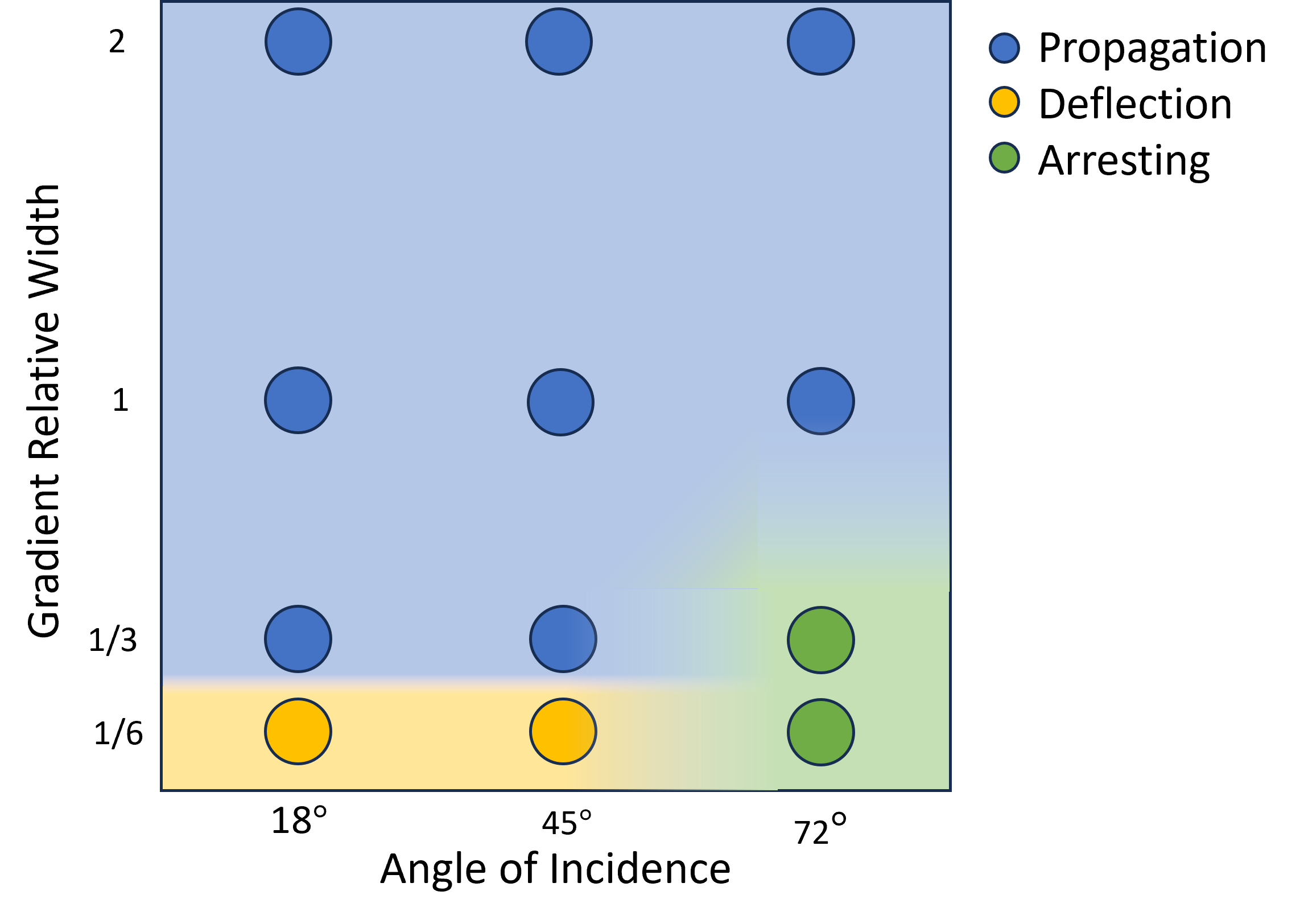}
          \caption{Diagram of crack behavior based on geometric conditions}
        \label{fig:ColoredGraph}
    \end{figure}

\begin{figure}[ht]
        \centering
        \begin{subfigure}{.3\textwidth}
          \centering
          \includegraphics[width=.95\linewidth]{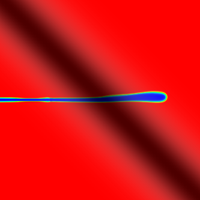}
          \caption{Case (1)}
          \label{fig:22}
        \end{subfigure}
        \begin{subfigure}{.3\textwidth}
          \centering
          \includegraphics[width=.95\linewidth]{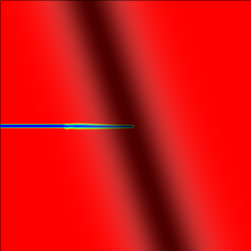}
          \caption{Case (2)}
          \label{fig:24}
        \end{subfigure}
        \begin{subfigure}{.3\textwidth}
          \centering
          \includegraphics[width=.95\linewidth]{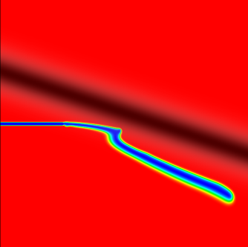}
          \caption{Case (3)}
          \label{fig:13}
        \end{subfigure}%
        \caption{Varying behavior in crack propagation across three simulations}
        \label{fig:PossibleBehavior}
    \end{figure}

        The grouping of these behaviors, and the conditions which lead to each crack propagation condition, is represented graphically in Figure \ref{fig:ColoredGraph}.
        There are three separated regimes for crack behavior, and these regimes are defined and separated only by material geometry.
        In our study of the parameter space, we found substantially more variance in crack propagation for narrow gradients. 
        Three representative cases of the behaviors in crack propagation are in Figure \ref{fig:PossibleBehavior}. 
        Figure \ref{fig:22} corresponds to $\theta=45^\circ$ gradient with $W/L = 1/3$, Figure \ref{fig:24} corresponds to a $\theta=72^\circ$ with $W/L = 1/3$, and Figure \ref{fig:13} corresponds to $\theta = 18^\circ$ with $W/L = 1/6$.
    
    \section{Conclusion} \label{Sec:Conclusion}
        In this work, we develop and demonstrate a phase field model for fracture in ductile functionally graded materials.
        Using this model, we identify specific crack propagation behaviors which vary between ductile and brittle functionally graded materials.
        We specifically focus on a gradient from a more brittle material to a tougher material, as cracks are more likely to initiate in a less tough material. 
        Our model highlights how the inclusion of plastic evolution and plastic dissipation fundamentally changes crack behavior compared to brittle models.

        We apply this model to determine potential fracture behaviors for ductile functionally graded materials, and determined three dominating fracture behaviors.
        These behaviors correspond to features of plastic strain, and, conversely, variance in plastic strain can be used to predict crack path. 
        The energy contribution of plastic strain directly corresponds to general crack behavior, and the resistance to accumulation of plastic strain is a defining driving factor in overall crack behavior.
        We have shown the relation between crack propagation and different material geometries, and determined the regions where each of these behaviors are seen.

        Collectively, this work demonstrates the importance of including plasticity in design of ductile FGMs. 
        While we demonstrate the importance of plasticity, we also acknowledge that in this work, wider gradients result in substantially reduced crack deflection. 
        Our study shows how laminate structures and narrow gradients, rather than wide FGMs, provide more promise for creating crack-deflecting behavior.
        We note that this behavior also depends on the choice of materials and does not necessarily negate the use of ductile FGMs for applications.
        While additional studies on mechanics of specific gradients should be conducted, our results indicate that common FGM uses, such as CTE matching, variance in ferromagnetism, and mechanical topological optimization, do not substantially impact bulk fracture behavior.
        In narrow gradients, however, our results indicate crack-deflecting or -arresting behavior can be engineered using a high rate of change in material properties in ductile FGMs. 
        
    \section*{CRediT authorship contribution statement}
    \textbf{Katherine Piper:} Conceptualization; Data curation; Formal analysis; Validation; Visualization; Writing - Original Draft.
    
    \textbf{Vinamra Agrawal:} Methodology; Formal analysis; Writing - review and editing; Supervision; Project Administration.
    
    \section*{Acknowledgments} 
        Authors acknowledge support for a portion of this work from Caltech Student Faculty Programs.
        Some simulations were performed using the Auburn University Easley high performance computing (HPC) cluster.
    \section*{Declaration of Competing Interests}
    The authors declare that they have no known competing financial interests or personal relationships that could have appeared to influence the work reported in this paper.

    \bibliographystyle{ieeetr}
    \bibliography{library.bib}
    
    \appendix

\end{document}